\documentclass[journal]{IEEEtran}
\usepackage{amsmath}
\usepackage{graphics}
\usepackage{epsfig}
\usepackage{amssymb}
\usepackage{t1enc}
\usepackage{multirow}
\usepackage{fancybox}
\usepackage{epic}
\usepackage{graphicx}
\usepackage{tikz}
\usepackage{booktabs}
\usepackage{pgfplots}
\usepackage{carl}
\usepackage[vlined,linesnumbered]{carlalgorithm}
\usetikzlibrary{patterns}

\newcommand{\setrandl}[0]{\stackrel{\mbox{\textnormal{\tiny $\$$}}}{\leftarrow}} %$ syntax highlighting fix

\newcommand{\url}[1]{{\it #1}}
\newcommand{\transpose}[0]{\textnormal{\tiny T}}
\newcommand{\scalar}[2]{\left\langle#1,#2\right\rangle}

\newcommand{\Ftwo}[0]{\mathbb{F}_2}
\newcommand{\textprob}[1]{\textnormal{\textsc{#1}}}
\newtheorem{prop}{Proposition}
\newtheorem{defn}{Definition}
\newtheorem{prblm}{Problem}

\newtheorem{lemma}{Lemma}
\newtheorem{thm}{Theorem}
\newtheorem{ex}{Example}

\ifCLASSINFOpdf
\else
\fi
\begin{document}
\title{A New Algorithm for Solving Ring-LPN with a Reducible Polynomial}

\author{Qian~Guo,
        Thomas~Johansson,~\IEEEmembership{Member,~IEEE,}
        and~Carl~L\"{o}ndahl% <-this % stops a space
\thanks{The authors are with the Department of Electrical and Information Technology, Lund University, Box 118, SE-22100 Lund, Sweden (e-mail: qian.guo@eit.lth.se; thomas.johansson@eit.lth.se; carl@grocid.net).}% <-this % stops a space
% <-this % stops a space
%\thanks{Manuscript received April 19, 2005; revised December 27, 2012.}
}

\maketitle

\begin{abstract} The \textprob{LPN} ({Learning Parity with Noise}) problem has recently proved to be of great importance in cryptology. A special and very useful case is the \textprob{Ring-LPN} problem, which typically provides improved efficiency in the constructed cryptographic primitive.
We present a new algorithm for solving the \textprob{Ring-LPN} problem
in the case when the polynomial used is reducible. It greatly
outperforms previous algorithms for solving this problem. Using the
algorithm, we can break the Lapin authentication protocol for the
proposed instance using a reducible polynomial, in about $2^{70}$ bit operations.
\end{abstract}


% \fi
\IEEEpeerreviewmaketitle

\section{Introduction}
\IEEEPARstart{L}{ight-weight} cryptography is a  field of cryptography inclined towards efficient cryptographic implementations, as a response to the demands when using highly constrained hardware in low-cost devices, such as passive RFID-tags and smart cards.

%People have designed many alternatives to address

There are trade-offs to consider, e.g., security, memory and
performance. Different constructions appear in different ends in the
trade-offs; for instance, AES and stream ciphers can be implemented
efficiently in hardware but do not offer \emph{provable security}.
Quite recently, a new trend arose in this area, building cryptographic
primitives from problems in learning theory.
Problems based on learning theory provide a complexity theoretical
foundation, on which the security of the cryptosystem can be based
upon. They also have the property of being easy and efficiently implemented, thereby making them appealing in light-weight cryptography.

\subsection{The LPN Problem}
Being a central problem in learning theory, the \textprob{LPN} problem
(\emph{Learning Parity with Noise}) has shown to be of significance in
the field of cryptography. It is a supposedly hard problem\footnote{\textprob{LPN} with adversarial errors is $\mathcal{NP}$-hard.}, and is not known to be susceptible to quantum
attacks, unlike some other classically hard problems such as factoring
and the discrete log problem.

%Efficient \textprob{LPN} algorithms would have impact on other areas as well. For instance, being able to solve the \textprob{LPN} problem efficiently would have strong implications on areas such as fast correlation attacks.

%Let $k$ be a security parameter and l

The problem can briefly be described as follows. Let $\vect s$ be a $k$-dimensional
binary vector. We receive a number $N$ of noisy versions of scalar
products of $\vect s$ from an oracle $\Pi_{\textnormal{LPN}}$, and our task is to recover $\vect  s$.

Let $\vect  y$ be a vector of length $N$ and let $y_i=\scalar{\vect
  s}{\vect r_i}$. For known random $k$ bit vectors $\vect r_1, \vect r_2, \ldots ,\vect
  r_N$, we can easily reconstruct an unknown ${\vect x}$ from $\vect
  y$ using linear algebra. In the LPN problem, however, we receive instead noisy versions
  of $y_i, 1\le i\le N$.

Writing the noise in position $i$ as $e_i,~ i=1,2,\ldots,N$, and
assuming each $\pr{e_i=1}$ to be small, we obtain
$$ v_i=y_i+e_i = \scalar{\vect x}{\vect r_i}+e_i .$$ In matrix form, the same is written as
$ \vect v =  \matr{A}\vect s    +\vect e,$
where $\vect v=\begin{bmatrix}
        v_1&v_2&\cdots&v_N
\end{bmatrix}$, and the matrix $\matr{A}$ is formed as
$\matr{A}=\begin{bmatrix}
        \vect r_1 & \vect r_2 & \cdots  & \vect r_N
\end{bmatrix}$.

\begin{table*}[!t]
        \begin{center}
                \caption{Comparison of different algorithms for attacking Lapin with reducible polynomial.}\label{table:ring-lpn-comparison}
                \begin{tabular}{@{}crrrrrrcr@{}}\toprule
                        Algorithm \phantom{~~} & & \multicolumn{3}{c}{Complexity ($\log_2$)} & & \\
                         \cmidrule{2-7}
                        & &\phantom{~~} Queries & \phantom{~~} Time  & \phantom{~~} Memory  & &  \\
                                \midrule
                        Levieil-Fouque \cite{LF06}  & &  82.0 & 103.4 & 100.6 &\\
                        Bernstein-Lange \cite{Bernstein} & &  79.3 & 102.9 & 97.9 &\\
                        Our attack (search)  & &  63 & 71.9 & 70.0 &\\ % 69.0
                        Our attack (decision)  & &  62 & 70.0 & 69.0 &\\ % 69.0 bits
                        \bottomrule
                \end{tabular}
        \end{center}
\end{table*}

\subsection{Constructions and Variations of LPN}
The LPN problem and its variations have been employed as the
underlying hard problem in a wide range of public-key cryptosystems, identification and authentication protocols, and zero-knowledge proofs.

The first actual usage of the LPN problem in cryptographic context can be traced back to 2001, when the Hopper-Blum (HB) identification protocol \cite{HB} was proposed. Being an intentionally minimalistic LPN based protocol, it was designed so that it could be executed by humans using only pen and paper. Much due to its simplicity, it is secure only in the \emph{passive} attack model. A couple of years later, Juels and Weis \cite{JW} along with Katz and Shin \cite{KatzShin} proposed a modified scheme, extending HB with one extra round. The modified scheme was named HB$^+$. Contrary to its predecessor, HB$^+$ was designed to be also secure in the \emph{active} attack model. However, it was discovered by Gilbert et al. \cite{Gilbert05} that the HB$^+$ protocol is susceptible to \emph{man-in-the-middle attacks} disproving active attack model security. The same authors \cite{Gilbert08} proposed later on another variation of the Hopper-Blum protocol called HB$^\#$, designed to resist their previous attack \cite{Gilbert05}. Apart from repairing the protocol, they solved the long-lived issue with large key-size or communication complexity by introducing the use of a slight variation of \textprob{LPN}, called \textprob{Toeplitz-LPN}. Although the use of \textprob{Toeplitz-LPN} has no documented weaknesses, its hardness remains unknown as of today.

During the time from when Hopper and Blum pioneered the use of \textprob{LPN} until today, a plethora of different proposals have hit the cryptographic society. Some of the most important ones are the proposals by Klitz et al. \cite{Klitz} and Dodis et al. \cite{Dodis} that showed how to construct message authentication codes based on \textprob{LPN}. The existence of MACs allows one to construct identification schemes that are provably secure against active attacks.

Most recently Heyse et al. \cite{Lapin} proposed a two-round identification protocol called Lapin. The protocol is based on \textprob{Ring-LPN} rather than \textprob{LPN}. Using the inherent properties of rings, the proposed protocol becomes very efficient and well-suited for use in constrained environments.
Briefly, in \textprob{Ring-LPN}, the oracle returns instead elements $v$, $v = s\cdot r + e$, from a polynomial ring $\gf{2}{}[x]/(f)$,
i.e., $v,s,r,e\in \gf{2}{}[x]/(f)$. The problem can use either an
irreducible polynomial $f$, of a reducible one. The choice of a
reducible polynomial can make use of the Chinese Remainder Theorem (CRT) to provide a very efficient implementation of the cryptographic primitive.

\subsection{Attacks}

\subsubsection{Attacks on Standard LPN}
The \textprob{LPN} problem can be viewed as a general decoding problem in coding theory. However, the usual
choice of parameters for the \textprob{LPN} problem is deviating from standard
parameters for a decoding problem. The \textprob{LPN} problem typically allows a
very large amount of oracle queries, say $N=2^{60}$ or even much larger; therefore, standard
algorithms for the general decoding problem, eg. information-set
decoding, will not always be very efficient
for such cases.

Instead, a slightly different type of algorithms have been suggested, among which we find
the  BKW algorithm \cite{BKW} proposed by Blum et al. It was later refined by Levieil
and Fouque \cite{LF06}. They gave better attacks by employing the Fast Walsh-Hadamard Transformation technique, and estimated the security level of \textprob{LPN} problems with different parameters.  Fossorier et al. \cite{Imai} suggested an algorithm which further improved the complexity by utilizing techniques from the area of fast correlation attacks. In the recent paper by Kirchner \cite{Kirchner}, it is proposed to exhaust the search space of the error, rather than the state. For a certain class of instances, i.e., when the error rate is low, Kirchner's technique greatly improves attack complexity.

Lyubashevsky's work \cite{Lyubashevsky} moved forward in the other research direction, i.e., when the number of required samples is bounded. Compared with BKW algorithm, he presented an asymptotically efficient algorithm with slightly increased time complexity, i.e., from $2^{\mathcal{O}(n/\log n)}$ to $2^{\mathcal{O}(n/\log \log n)}$, while obtaining much query efficiency.

\subsubsection{Attacks on Ring-LPN}
As the \textprob{Ring-LPN} instances are standard \textprob{LPN} instances, the attacking algorithm for the latter one is applicable to its ring instance. The pioneering researchers (eg. Lapin \cite{Lapin}) used the hardness level of the \textprob{LPN} problem obtained from \cite{LF06} to measure the security of their authentication protocol based on the \textprob{Ring-LPN} problem. Almost at the same time, Bernstein and Lange \cite{Bernstein} realized  that simply ignoring the ring structure is inappropriate since this special algebraic property may reveal information about the secret, and subsequently derived an improved attack taking advantage of both the ring structure and Kirchner's technique. Their attack is generic since it applies to \textprob{Ring-LPN} implemented with both reducible and irreducible polynomials, and is advantageous in the sense of memory costs as well as query complexity. However, even for the time-optimized case\footnote{We found that with parameters $q=2^{58.59}$, $a=6$, $b=65$, $l=53$ and $W=4$, the complexity is $2^{85.9}$. The time complexity is slightly lower than stated in \cite{Bernstein}.}, with around $2^{81}$ bits of memory, it requires quite a large number of bit operations, i.e., about $2^{88}$, far away from breaking the $80$-bit security of Lapin protocol.

\subsection{Our Contribution}

We propose a new algorithm to solve the reducible case of \textprob{Ring-LPN}. By investigating more on the properties of the ring structure and the reducibility of the polynomial, we demonstrate that if the minimum weight of the linear code defined by the CRT transform is low, then the problem is effortless to solve, hence providing a design criteria for cryptosystems based on the hardness of \textprob{Ring-LPN} with a reducible polynomial.

We then specify the attack for Lapin \cite{Lapin} and obtain a complexity gain that makes it possible to break the claimed 80-bit security. In Table \ref{table:ring-lpn-comparison}, we compare the
complexity of our algorithm with the best known algorithms\footnote{We choose parameters to optimize their time complexity.} designed to solve \textprob{LPN} and \textprob{Ring-LPN}. The time complexity is measured in bit operations and memory complexity is measured in bits.

The organization of the paper is as follows. In Section
\ref{ring-lpn:the-problem}, we give some preliminaries and introduce the
\textprob{Ring-LPN} problem in detail. We describe our new generic attack in
Section \ref{ring-lpn:the-attack} and then present a special version that is efficient for the proposed reducible instance of Lapin in Section \ref{ring-lpn:Improved-version}.
In Section
\ref{ring-lpn:the-analysis} we analyze its complexity. The numerical results
when the algorithm is applied on Lapin are given in Section \ref{ring-lpn:the-results} and Section \ref{ring-lpn:the-conclusion} concludes the paper.

\section{The Ring-LPN problem}\label{ring-lpn:the-problem}

\subsection{Polynomials and Rings}
Consider a polynomial $f(x)$ over $\Ftwo$ (simply denoted $f$). The degree of $f$ is denoted by
$\deg f$. For any two polynomials $f, g$ in the quotient ring $\Ftwo [x]$, long division of
polynomials tells us that there are unique polynomials $q,r$ such that
$g=qf+r$. The unique polynomial $r \in \gf{2}{}[x]$ with $\deg r < \deg f$
is the representative of $g$ in the quotient ring $\gf{2}{}[x]/(f)$ and is
denoted $g \bmod f$. We define $R$ to be the quotient ring
$\gf{2}{}[x]/(f)$.  So $R$ consists of all polynomials in $ \gf{2}{}[x]$ of degree
less than $\deg f$ and arithmetics are done modulo $f$.

If the polynomial $f$ factors such that every factor is of degree
strictly less than $f$, then the polynomial $f$ is said to be
\emph{reducible}; otherwise, it is said to be \emph{irreducible}. When
a reducible polynomial factors as $f=f_1 f_2  \cdots f_m$, where $f_i$ is relatively prime to $f_j$ for $1 \leq i,j \leq m$ and $i\neq j$, there exists a unique representation for every $r \in R$ according to the Chinese Remainder Theorem, i.e.,
$$
    r \mapsto (r \bmod {f_1},r \bmod {f_2},\ldots ,r\bmod {f_m}).
$$

\subsection{Distributions}

Let $\distr{Ber}_\eta$ be the Bernoulli distribution and $X$ be a
binary random variable. It is said that $X$ is distributed according
to the Bernoulli distribution with parameter $\eta$, if $\pr{X = 1} =
\eta$ and $\pr{X = 0} = 1 - \pr{X = 1} = 1-\eta$. Then, the bias
$\epsilon$ of $X$ is given by $\pr{X = 0} =
\frac{1}{2}\p{1+\epsilon}$, i.e. $\epsilon=1-2\eta$ . Let $r \setrandl\textnormal{\textsf{Ber}}_\eta^R$ denote that the coefficients of the ring element $r \in R$ are drawn randomly according to the distribution $\textnormal{\distr{Ber}}_\eta$.
The uniform distribution is denoted $\distr{U}$. Whenever we draw an element uniformly from $R$, we denote it $r \setrandl \distr{U}^{R}$.

\subsection{Formal Definition of Ring-LPN}
Being a subclass of \textprob{LPN}, the \textprob{Ring-LPN} problem is
defined similarly. Fix an unknown value $s\in R$, where $s \setrandl \distr{U}^{R}$.
We can request samples depending on $s$ through an oracle, which we define as follows.
\begin{defn}[\textprob{Ring-LPN} oracle] A \textprob{Ring-LPN} oracle $\Pi_{\textprob{Ring-LPN}}^\epsilon$ for an unknown polynomial $s \in R$ with $\eta \in (0,\frac{1}{2})$ returns pairs of the form \[\p{r ,~ r \cdot s + e},\]
        where $r \setrandl \distr{U}^{R}$ and $  e \setrandl \textnormal{\distr{Ber}}_\eta^R$.
        An extreme case is, when $\eta$ is exactly $\frac{1}{2}$, that the oracle $\Pi_{\textprob{Ring-LPN}}^0$ outputs a random sample distributed uniformly on $R\times R$.
\end{defn}

The problem is now to recover the unknown value $s$ after a number $q$ of queries to the oracle.
 We define the search problem version of \textprob{Ring-LPN} in the following way.
\begin{prblm}[\textprob{Ring-LPN}] The search problem of
 \textprob{Ring-LPN} is said to be $(t,q,\delta)$-solvable if there
 exists an algorithm $\mathcal{A}(\Pi_{\textprob{Ring-LPN}}^\epsilon)$
 that can find the unknown polynomial $s \in R$ in time at most $t$
 and using at most $q$ oracles queries such that
\[ \pr{  \mathcal{A}(\Pi_{\textprob{Ring-LPN}}^\epsilon) = s} \geq \delta.\]
\end{prblm}
The decisional \textprob{Ring-LPN} assumption, states that it is hard
to distinguish uniformly random samples from pairs from the oracle
$\Pi_{\textprob{Ring-LPN}}^\epsilon$. It can be expressed as follows.

\begin{prblm}[Decisional \textprob{Ring-LPN}] The decision problem of \textprob{Ring-LPN} is said to be $(t,q,\delta)$-solvable if
there exists an algorithm  $\mathcal{D}$ such that
$$\left| \pr{  \mathcal{D}(\Pi_{\textprob{Ring-LPN}}^\epsilon) = \textnormal{\textsf{yes}}} - \pr{\mathcal{D}(\Pi_{\textprob{Ring-LPN}}^0) = \textnormal{\textsf{yes}}}\right| \geq \delta,$$
and $\mathcal{D}$ is running in time $t$ and making $q$ queries.
\end{prblm}

The hardness of \textprob{Ring-LPN} is unknown, but the \textprob{LPN} problem has been shown to be $\mathcal{NP}$-hard in the worst-case. The assumption is that \textprob{Ring-LPN} is also hard.

In the paper by Heyse et al. \cite{Lapin}, it was proposed to use
\textprob{Ring-LPN} as the underlying hard problem to build an
authentication protocol. The security relies on the assumption that
\textprob{Ring-LPN} is as hard as \textprob{LPN}. However, this is a
conjecture as there is only a reduction from \textprob{Ring-LPN}
to \textprob{LPN}, but not the converse. In the following, we show how to reduce \textprob{Ring-LPN} to \textprob{LPN}.

\subsubsection{Transforming {Ring-LPN} to {LPN}}\label{SubsectionTransMapping}
Given the polynomial $r$ with $\deg r = t$, we denote $\vect r$ as its
coefficient vector,  i.e., if $r$ equals $ \sum_{i=0}^{t}r_ix^i$, then
$\vect r = \begin{bmatrix} r_0 & r_1 & \cdots & r_{t}\end{bmatrix}$.
With this notation,  we can define a mapping from one \textprob{Ring-LPN} instance to $d$ standard \textprob{LPN} instances represented in the following matrix form:
\begin{eqnarray}
  \tau:~~~~~~ R \times R  & \rightarrow & \Ftwo^{t\times t} \times \Ftwo^t \nonumber \\
        (r,r \cdot s + e)   & \mapsto   &        (\matr A, \matr A  \vect s + \vect e)
\end{eqnarray}
where the $i$-th column of the matrix $\matr A$ is the transposed coefficient vector of $r \cdot x^i \bmod f$.

\subsubsection{ {Ring-LPN} with a Reducible $f$}
In this paper we consider the \textprob{Ring-LPN} problem with the
ring $R=\gf{2}{}[x]/(f)$, where the polynomial $f$ factors into
distinct irreducible factors over $\gf{2}{}$. That is, the polynomial
$f$ is written as $f = f_1f_2 \cdots f_m$, where each $f_i$ is
irreducible. One of the specified instances of Lapin \cite{Lapin} uses
a product of five different irreducible polynomials. This instance is
the main target for cryptanalysis in this paper.

\subsection{Basics on Coding Theory}
For later use, we end this section by reviewing some basics on
\emph{linear codes} over the binary field. A linear code $\ccode$ is a
$k$-dimensional subspace of an $n$-dimensional vector space. The
elements of $\ccode$ are called \emph{codewords} and the Hamming
weight of a codeword is defined as number of non-zero entries. The
\emph{minimum distance} of a linear code is defined as the lowest
weight among the weights of all codewords.

\begin{defn}[Generator matrix]
    A generator matrix $\matr G$ for $\ccode$ is defined as a $k \times n$ matrix whose span is the codeword space $\ccode$, i.e.,
    $$ \ccode = \set{\vect x \matr G }{\vect x \in \gf{2}{k}}.$$
\end{defn}

The existence of codes with specific parameters is sometimes
guaranteed through the famous GV bound.
\begin{thm}[Gilbert-Varshamov bound] Let $n$, $k$ and $d$ be positive integers such that
$$ \textnormal{Vol}(n-1,d-2) < 2^{n-k},$$
where $\textnormal{Vol}(n-1,d-2)=\sum_{i=0}^{d-2} {n-1 \choose i}$.
   Then, there exists an $[n,k]$ linear code having minimum distance at least $d$.
\end{thm}

A {\em random} linear code is a code $\ccode$, where the entries of the
generator matrix $\matr G$ has been selected according to an
i.i.d. uniform distribution. It is known that the minimum distance $d$
for a random linear code asymptotically follows the GV bound.

\section{The New Algorithm for Ring-LPN with Reducible Polynomial}\label{ring-lpn:the-attack}
The purpose of the paper is to describe a new algorithm (Algorithm \ref{alg:ring-lpn-attack}) for the
\textprob{Ring-LPN} problem with a reducible polynomial. We describe the
algorithm as follows. First, we reduce the problem into
a smaller one, while keeping the error at a reasonable level. Then, we
further reduce the unknown variables using well-established collision
techniques. The last step consists of exhausting the remaining unknown variables in an efficient way.

\subsection{A Low-Weight Code from the CRT Map}

%In the reducible case of \textprob{Ring-LPN}, the ring elements
Let $f = f_1 f_2  \cdots f_m$ be a reducible polynomial. In the following, the underlying irreducible polynomial $f_i$ in the quotient ring $\gf{2}{}[x]/(f_i)$ is fixed. Therefore, w.l.o.g., we denote it as $f_1$ for notational simplicity. Let $\deg f = t$ and $\deg f_1 = l$.

% The following applies to any $f_i$, $1 \leq i \leq m$, but for notational simplicity we denote the polynomial of the smaller ring $f_1$.

\begin{prop}\label{prop:ring-lpn-crt-transform-to-generator-matrix}
There exists a (surjective) linear map from $\psi:R \rightarrow \gf{2}{}[x]/(f_1)$ determined by the CRT transform. The linear map can be described as a $[t,l]$ linear code\footnote{The code is a punctured LFSR code.} with generator matrix $G_\psi$, in which the $i$-th column is the transposed coefficient vector of the polynomial $x^{i-1}\bmod{f_1}$.
\end{prop}

More specifically, a received sample $(r,r\cdot s+e)$ from the
\textprob{Ring-LPN} oracle $\Pi_{\textprob{Ring-LPN}}^\epsilon$ can be
transformed into a considerably smaller instance {\small $$\psi:(r,r\cdot s+e)
\mapsto (r \bmod{f_1},(r\bmod{f_1})\cdot (s\bmod{f_1})+e\bmod{f_1}).$$}
For simplicity, we  write $\hat{r}=r
\bmod{f_1}$, $\hat{s}=s \bmod{f_1}$ and $\hat{e}=e \bmod{f_1}$. As before, we may write $\hat{r} = \sum_{i = 0}^{l-1}\hat{r}_i x^i$, etc.

The new instance has a smaller dimension, as $\deg f_1 < \deg f$. However, the distribution of the errors is also changed. The error distribution in the larger ring is $\textsf{Ber}_{\frac{1}{2}(1-\epsilon)}$, but in the smaller ring each noise variable $(\hat{e}=e \bmod{f_1})$ is a sum of several entries from $e$. The number of noise variables that constitutes a new error position $(\hat{e}=e \bmod{f_1})$ depends entirely on the relation between $f$ and $f_1$.

%The following is an actual example of a proposed construction given in \cite{Lapin}:

The following is an example that chooses $f_1$ to be one of the irreducible polynomials employed in \cite{Lapin}.

\begin{ex}\label{ex:ring-lpn-low-weight-relation} Let $\deg{f}=621$,
  let $f_1 = x^{127}+x^8+x^7+x^3+1$ and assume $f_1|f$. %Then the
%  constant term in an element $\hat{r}$ can be expressed as a sum of
%  terms of the form $x^q \bmod{f}$, e.g., the following equation holds:
%$$
%\begin{array}{rll}
%    1 \bmod{f_1} =
%    & 1+&x^{127}+x^{246}+x^{247}+x^{251}+x^{254}+x^{365}+\\
%    && x^{367}+x^{375}+x^{381}+x^{484}+x^{485}+x^{486}+\\
%    && x^{487}+x^{489}+x^{491}+x^{492}+x^{495}+x^{499}+\\
%    && x^{500}+x^{501}+x^{502}+x^{505}+x^{508}+x^{603}+x^{607}.
%\end{array}
%$$
If we consider $\hat{e}_i$ to be the $i$-th entity of the coefficient
vector of  $\hat{e}= e \bmod {f_1}$, then we express $\hat{e}_0$ as a
sum of bits from $e$ as follows,
$$
\begin{array}{rll}
   \hat{e}_0 =
    & e_0+&e_{127}+e_{246}+e_{247}+e_{251}+e_{254}+e_{365}+\\
    && e_{367}+e_{375}+e_{381}+e_{484}+e_{485}+e_{486}+\\
    && e_{487}+e_{489}+e_{491}+e_{492}+e_{495}+e_{499}+\\
    && e_{500}+e_{501}+e_{502}+e_{505}+e_{508}+e_{603}+e_{607}.
\end{array}
$$

\end{ex}

In particular, we are interested in linear relations that have as few
noise variables from $e$ involved as possible. We use the Piling-up
lemma to determine the new bias in $\hat{e}$ after summing up a number of error bits.

\begin{lemma}[Piling-up lemma]
     Let $X_1,X_2,...X_n$ be i.i.d binary variables where each $\pr{X_i = 0} = \frac{1}{2}( 1 + \epsilon_i)$, for $1 \leq i \leq n$. Then,
    %$$ \pr{X_1 + X_2 + \cdots + X_n = 0} = 1/2 + 2^{n-1}\prod_{i=1}^n\epsilon_i,$$
    %and thus,
    $$ \pr{X_1 + X_2 + \cdots + X_n = 0} = \frac{1}{2}\p{1+\prod_{i=1}^n\epsilon_i}.$$
\end{lemma}

For instance, the linear relation given in Example
\ref{ex:ring-lpn-low-weight-relation} has weight 26. Hence, by the
Piling-up lemma, the bias of $\hat{e}$ in that particular position (position 0) is $ \epsilon^{26}$, i.e., $\pr{\hat{e}_0 = 1} = \frac{1}{2}(1-\epsilon^{26})$.

\vspace{3ex}

\textbf{Note:} We assume that $f_1$ is an irreducible polynomial throughout the paper. However, this condition is not a necessity, as the essential feature of the new attack is a CRT map. Actually, it is sufficient if the two polynomials $f_1$ and $f/f_1$ are coprime; for example, we could set the polynomial with small degree to be the product of several irreducible polynomials and obtain a solution as well.

\subsection{Using Low-Weight Relations to Build a Distinguisher}
%If the \textprob{Ring-LPN} oracle $\Pi_{\textprob{Ring-LPN}}^\epsilon$
%is distinguishable from a random oracle, it has non-random properties
%that can be exploited. Hopefully, these relations  can also be used by
%an adversary to find the secret polynomial. We will now show how to
%build such a distinguisher for \textprob{Ring-LPN} with a reducible
%polynomial using the previously presented ideas. Again, consider $f_i$ fixed.
%
%\begin{prop}[Lower-bound on required samples]
%    The smallest number of error variables that can act upon a position (or sum of positions) in the smaller ring is given by the minimum distance $d_\psi$ of the linear code $\matr{G}_\psi$.
%\end{prop}
%
%As explained, sparser codewords yield larger biases, thereby allowing more efficient distinguishing attacks. Therefore, to build an efficient distinguisher $\mathcal{D}$ we first need to find low-weight codewords in $G_\psi$. For this, we can use an information-set decoding algorithm. However, in the cases we have observed, we suspect that the sparsest codewords actually are among the rows of $\matr{G}_\psi$ as their weights are below the GV-bound. Such investigation on searching for codewords with lowest weight would be of greater interest when $f_i$ is a dense polynomial.

We will now show how to build a distinguisher for \textprob{Ring-LPN} with a reducible polynomial using the CRT transformation described in the previous section.

In Example \ref{ex:ring-lpn-low-weight-relation}, we give a linear relation expressing an error variable $\hat e_0$ in the smaller ring as a sum of relatively few error variables in the larger ring. In our example, the polynomial $f_1$ >>behaves well<<; it is very sparse and yields a low-weight relation expressing a single noise variable $\hat e_0$. However, this will generally not be the case: it may be very difficult to find a desired linear relation with few error variables.

We observe an interesting connection between this problem and searching for codewords with minimum distance in a linear code. The CRT transformation can be viewed as
$$
\begin{bmatrix}
    1 \bmod{f_1}\\
    x\bmod{f_1}\\
    \vdots\\
    x^{t-1}\bmod{f_1}
\end{bmatrix}=
\underbrace{
\begin{bmatrix}
    \vect g_0 \\
    \vect g_1\\
    \vdots \\
  \vect g_{l-1} \\
    \hline
     \vect g_l \\
    \vect g_{l+1}\\
    \vdots \\
  \vect g_{t-1} \\
\end{bmatrix}}_{=G_\psi^\transpose}
\begin{bmatrix}
    1 \\
    x\\
    \vdots\\
    x^{l-1}
\end{bmatrix},
$$
where the top part of $G_\psi^\transpose$ is an identity matrix and each row $\vect g_i$, for $0\leq i \leq t-1$, is the coefficient vector of the polynomial $x^{i}\bmod{f_1}$.

Thereby, expressing the error polynomial in the smaller ring

\begin{eqnarray*}
                e
\bmod f_1 &=& \underbrace{\begin{bmatrix}e_0 & e_1 & \cdots e_{t-1}\end{bmatrix}}_{\defeq \vect e^\transpose}
G_\psi^\transpose
\begin{bmatrix}
    1 \\
    x\\
    \vdots\\
    x^{l-1}
\end{bmatrix} \\
                 &=& \underbrace{\begin{bmatrix}
\hat e_0 & \hat e_1 & \cdots \hat e_{l-1}
\end{bmatrix}}_{\defeq \hat{\vect e}^\transpose}
\begin{bmatrix}
    1 \\
    x\\
    \vdots\\
    x^{l-1}
\end{bmatrix},
              \end{eqnarray*}
we obtain $$\vect e^\transpose G_\psi^\transpose = \hat{\vect e}^\transpose \Rightarrow  \hat{\vect e} =  G_\psi \vect e.
$$

%\begin{eqnarray*}
%              % \nonumber to remove numbering (before each equation)
%                e
%\bmod f_1 &=& \underbrace{\begin{bmatrix}e_0 & e_1 & \cdots e_{t-1}\end{bmatrix}}_{\defeq \vect e^\transpose}
%\begin{bmatrix}
%    1 & 0 & 0  & \cdots \\
%    0 & 1 & 0 & \cdots\\
%    \vdots & & \ddots & \\
%\end{bmatrix}
%\begin{bmatrix}
%    1 \\
%    x\\
%    \vdots\\
%    x^{l-1}
%\end{bmatrix} \\
%                 &=& \underbrace{\begin{bmatrix}
%\hat e_0 & \hat e_1 & \cdots \hat e_{l-1}
%\end{bmatrix}}_{\defeq \hat{\vect e}^\transpose}
%\begin{bmatrix}
%    1 \\
%    x\\
%    \vdots\\
%    x^{l-1}
%\end{bmatrix},
%              \end{eqnarray*}

Let $d_\psi$ be the minimum distance of the linear code generated by $\matr G_\psi$. Then by definition, there exists at least one vector $\vect m$ such that the product $\vect m \matr G_\psi$ has Hamming weight exactly $d_\psi$. More specifically,
$$  \scalar{\hat{\vect e}}{ \vect m } =  \scalar{\vect e } {\vect m \matr G_\psi}, $$
where $\scalar{\vect e}{ \vect m \matr G_\psi}$ is a sum of $d_\psi$ noise variables. Thus, according to Piling-up lemma we obtain the following proposition.

\begin{prop}[Estimate of required samples]\label{prop:estimation-of-samples}
    If the minimum distance of the linear code $\matr G_\psi$ is $d_\psi$, then the largest bias of some linear combination of noise variables in the smaller ring is no less than $\epsilon^{d_\psi}$.
\end{prop}

Consequently, in order to determine the security related to a certain polynomial, we need to determine the minimum distance of the code generated by $\matr G_\psi$. By applying well-known algorithms such as \emph{information-set decoding} (ISD) algorithms e.g. \cite{Stern89}, we can find the minimum distance.

In the example of Lapin, applying ISD algorithms is not necessary since the polynomials $f_i$, $1 \leq i \leq 5$ are very sparse and admit a very low row-weight of the generator matrix, well below the GV-bound (which is around 154).

\subsection{Recovering the Secret Polynomial}
%We have used $\hat e$ to denote the an element $e$ in the smaller ring. Since most operation will occur in the smaller ring, we will now use $e$ to denote the element in the smaller ring.
After determining the strongest bias ($ \scalar{\hat{\vect e}}{\vect m}$), we move to the recovery part.

\subsubsection{Transforming}

 Recall that $\deg f_1 = l$. We ask the oracle $\Pi_{\textprob{Ring-LPN}}^\epsilon$ for $N$ samples $(\hat{r}_{(i)},\hat{s}\cdot \hat{r}_{(i)}+\hat{e}_{(i)})$ and then convert each of them to $l$ standard \textprob{LPN} samples by the mapping $\tau$ defined in Section \ref{SubsectionTransMapping}. Write these samples in the matrix form $(\matr{\hat A_i}, \matr{\hat A_i}\hat{\vect s} + \hat{\vect e}_i)$. Then, multiplying with vector $\vect m$, we construct a new $\textprob{LPN}$ sample,
$$(\vect m \matr{\hat A_i}, \vect m \matr{\hat A_i}\hat{\vect s} + \scalar{ \hat{\vect e}_i}{\vect m} ),$$
from each $\textprob{Ring-LPN}$ sample. According to Proposition \ref{prop:estimation-of-samples}, the created samples are with large bias, i.e., no less than $\epsilon^{d_\psi}$.

The overall computational complexity of this step is bounded by,
$$C_1 = Nl^2 + Nl\cdot (l+1) = Nl\cdot (2l + 1).$$

\subsubsection{Birthday}
Put these new samples in a data structure that can be accessed in constant time (e.g., a hash table), indexed by its last $k$ entries of the vector $\vect m \matr{\hat A_i}$.  Then
a collision between vectors $\vect m \matr{\hat A_i}$ and $\vect m \matr{\hat A_j}$ (denoted $\hat{\hat{\vect r}}_i$ and $\hat{\hat{\vect r}}_j$, respectively),
$$(\hat{\hat{\vect r}}_i, \hat{\hat{\vect r}}_i\cdot\hat{\vect s} + \scalar{ \hat{\vect e}_i}{\vect m} )+(\hat{\hat{\vect r}}_j, \hat{\hat{\vect r}}_j\cdot \hat{\vect s} + \scalar{ \hat{\vect e}_j}{\vect m} ) = (\hat{\vect r}',\hat{\vect r}' \cdot \hat{\vect s} + \hat{ e}'), $$
yields a vector $\hat{\vect r}'$ that has at most $l - k$ nonzero positions. The number of such samples is approximately $M = N^2 /2^k$. The new samples, such as $(\hat{\vect r}',\hat{ v}')$, depend only on $l-k$ coefficients of the secret $\hat{s}$ and has a bias that is $$\epsilon' = \epsilon^{2d_\psi}.$$ Calculating the divergence between the distribution of error in $(\hat{\vect r}',\hat{ v}')$ and the uniform distribution, we find that the number of samples required is $M \geq 1/{\epsilon^{4d_\psi}}$.

Storing the $N$ \textprob{LPN} samples uses $lN$ bit-operations, and performing the birthday procedure requires $lN^2/2^k$ bit-operations. Thus, the total complexity of this step is,
$$C_2 = Nl\cdot ( 1 + \frac{N}{2^k}).$$

Thus, at this point, we have generated $M$ vector samples $(\hat{\vect
r}'_i, \hat{v}_i)$. All $\hat{\vect r}'_i$ vectors have dimension no more than $l-k$
as we cancelled out $k$ bits of $\hat{\vect r}'_i$. Hence, it is enough to
consider only $l-k$ bits of $\hat{s}$, i.e., we assume that $\hat{s}$ is of dimension $l-k$.  We are then prepared for the final step.

\subsubsection{Distinguishing the Best Candidate}
Group the samples $(\hat{\vect r}'_i, \hat{v}_i)$ in sets $L(\hat{\vect r}_i')$ according to $\hat{\vect r}'_i$ and then define the function $f_L(\hat{\vect r}'_i)$ as
$$f_L(\hat{\vect r}_i') = \sum_{(\hat{\vect r}'_i, \hat{v}_i)\in L(\hat{\vect r}_i')}(-1)^{\hat{v}_i}.$$
The Walsh transform of $f_L$ is defined as
$$F(\hat{\vect s})= \sum_{\hat{\vect r}'_i} f_L(\hat{\vect r}'_i)(-1)^{\scalar{\hat{\vect
  s}}{\hat{\vect r}'_i}}.$$
Here we exhaust all the $2^{l-k}$ candidates of $\vect s$ by
  computing the Walsh transform.

Given the candidate $\hat{\vect s}$, $F(\hat{\vect s})$ is the difference between
the number of predicted $0$ and the number of predicted
$1$ for the bit $\hat{v}'_i + \scalar{\hat{\vect
  s}}{\hat{\vect r}'_i}$. If $\hat{\vect s}$ is the correct guess, then it is distributed
according to
$\textnormal{\distr{Ber}}_{\frac{1}{2}(1-\epsilon^{2d_\psi})}$; otherwise,
it is considered random. Thus, the best candidate $\hat{\vect s}_0$ is the
one that maximizes the absolute value of $F(\hat{\vect s})$, i.e. $\hat{\vect s}_0
= \arg \max_{\hat{\vect s}\in F_2^{l-k}}|F(\hat{\vect s})|$, and we need
approximately $\epsilon^{4d_\psi}$ samples to distinguish these two
cases. Note that false positives are quickly detected in an additional
step and this does not significantly increase complexity.

We employ Fast Walsh-Hadamard Transform technique to accelerate the distinguishing step. For well-chosen parameters, the complexity is approximately,
$$C_3 = (l-k)2^{l-k}.$$

\algnoinput{Partial recovery of \textprob{Ring-LPN}}{alg:ring-lpn-attack}
        {
                \begin{algorithm*}[!t]
                \dontprintsemicolon
                \;
        (\textbf{{Preprocessing}}) Determine the minimum weight $d_\psi$ of the linear code generated by the CRT transformation and find its corresponding linear relation $\vect m$. \;
         Ask the oracle $\Pi_{\textprob{Ring-LPN}}^\epsilon$ for $N$ samples, and then transform each of them to a standard $\textprob{LPN}$ sample with the largest bias, which is no less than $\epsilon^{d_\psi}$.\;

        Use the Birthday technique to reduce the secret length at the cost of decreasing the bias.\;
        Perform Fast Walsh-Hadamard Transform on the remaining $l-k$ bits of $\hat{\vect s}$.\;
        Output the $\hat{\vect s}_0$ that maximizes the absolute value of the transform.\;
                \end{algorithm*}
        }

\vspace{3ex}

From the new algorithm, there are some important consequences to
consider when choosing parameters to thwart our attack. We give some very brief comments, assuming that every smaller ring is of approximately the same size:
\begin{enumerate}
\item Choosing a large number of factors in $f$ seems a bit dangerous,
  as the dimension of the code $G_\psi$ becomes small. In our attack
  we used a birthday argument to reduce the dimension of $\hat{s}$,
  but this might not be necessary if the dimension is already very
  low. Instead we may search for special $\hat{r}$ (e.g. many
  $\hat{r}=1$) values that allows quick recovery of $\hat{s}$.

\item One should use irreducible polynomials $f_i$ with degree around
  $\frac{n}{m}$ for $1 \leq i \leq m$ such that for every $f_i$ the
  corresponding linear code $G_\psi$ has minimum distance as large as
  possible. From the GV bound, we know roughly what to expect.
  However, the GV-bound may not be asymptotically true for codes generated by the CRT-transform.

\item Following this line, a necessary but probably insufficient
  condition on $\epsilon$ is that $1/\epsilon^{4d_\psi} \geq 2^{b}$
  for $b$-bit security\footnote{This is just one concern as there may
  be many other aspects to be taken into consideration that will make
  the problem solvable. For instance, if $d_\psi$ and $\epsilon$ are
  very large the constraint can be satisfied while the problem in a
  larger ring remains easy.}.

\end{enumerate}

\section{The Improved Version for the Proposed Instance of Lapin}\label{ring-lpn:Improved-version}
In \cite{Lapin}, Heyse et al. employ the following reducible polynomial,
{\small \begin{equation*}
    \begin{array}{rl}
     f = & \underbrace{\p{x^{127}+x^8+x^7+x^3+1}}_{f_1}\cdot \underbrace{\p{x^{126}+x^9+x^6+x^5+1}}_{f_2}\\
     & \cdot \underbrace{\p{x^{125}+x^9+x^7+x^4+1}}_{f_3}\cdot \underbrace{\p{x^{122}+x^7+x^4+x^3+1}}_{f_4}\\
     & \cdot \underbrace{\p{x^{121}+x^8+x^5+x^1+1}}_{f_5},
     \end{array}
\end{equation*}}
as the underlying structure of the quotient ring $R$. This parameter setting is then adopted in a more recent paper \cite{Mask-Lapin} by Gaspar at al., to show that the original protocol and its hardware variant, Mask-Lapin, have much gain, compared with the implementation from block ciphers (e.g., AES), in the sense of resisting power analysis attacks by masking.

However, in the sense of thwarting the new attack of this paper, it is not a good selection. We give two reasons as follows.
\begin{itemize}
  \item As stated previously, for any polynomial $f_i$ ($1\leq i \leq 5$), the generator matrix $G_{\psi}$ of the corresponding code is sparse, hence yielding that we could roughly adopt one row vector in $G_{\psi}$ as its minimum weight codeword. Then, the vector $\vect m$ is of Hamming weight $1$ and we could save the computational cost for linearly combining several samples to form a new one with the largest bias.
  \item Secondly, for the Lapin instance, the largest bias always holds at the last row of the generator matrix $G_{\psi}$, when modulo operation is taken over each irreducible factor. Furthermore, for \textprob{Ring-LPN} samples $(\hat{r}_{(i)},\hat{s}\cdot \hat{r}_{(i)}+\hat{e}_{(i)})$, if we find collisions on the last $k$ positions of $\hat{r}_{(i)}$ ($k$ is larger than $10$), then the last row vector in the matrix form of the merged sample $(\hat{r}',\hat{r}' \cdot \hat{s} + \hat{e}')$ is of the form
      \begin{equation}\label{eq:transform-last-row}
      \begin{bmatrix}
      0 & \ldots & 0 & \hat{r}'_{l-k-1} & \hat{r}'_{l-k-2} & \cdots & \hat{r}'_{0}
      \end{bmatrix}.
      \end{equation}
      This vector can be read from the polynomial $\hat{r}'$ directly, without any computation cost.
\end{itemize}

Therefore, we could present a specific algorithm for solving the Lapin instance, see Algorithm \ref{alg:improved-attack}, which is more efficient than the generic one. After determining the weight $w$ of the last row vector in the generator matrix $G_{\psi}$, we ask the oracle $\Pi_{\textprob{Ring-LPN}}^\epsilon$ for $N$ samples $(\hat{r}_{(i)},\hat{s}\cdot \hat{r}_{(i)}+\hat{e}_{(i)})$ and search for collisions by the last $k$ coefficients of $\hat{r}$ directly. Then, for each collision represented by a merged \textprob{Ring-LPN} sample $(\hat{ r}',\hat{ v}')$, we construct a new standard  \textprob{LPN} sample, where the vector is generated by (\ref{eq:transform-last-row}), and the observed value is the coefficient of $x^{l-1}$ in the polynomial $\hat{ v}'$. These samples are with the largest bias.
The distinguishing step is the same as that in the generic algorithm and we present the detailed complexity analysis and numerical results in the consecutive sections.

\algnoinput{Improved partial key recovery for Lapin}{alg:improved-attack}
        {
                \begin{algorithm*}
                \dontprintsemicolon
                \;
        (\textbf{{Preprocessing}}) Find the weight $w$ of the $(l-1)$-th row in the generator matrix of the CRT transformation. \;
        Ask the oracle $\Pi_{\textprob{Ring-LPN}}^\epsilon$ for $N$ samples, index them by the last $k$ coefficients of $\hat{r}$, and then search for all collisions $(\hat{r}',\hat{r}' \cdot \hat{s} + \hat{e}_1 + \hat{e}_2)$ $ = (\hat{r}_1,\hat{s}\cdot \hat{r}_1+\hat{e}_1)+(\hat{r}_2,\hat{s}\cdot \hat{r}_2+\hat{e}_2)$, where the last $k$ coefficients of $\hat{r}_1$ and $\hat{r}_2$ are the same.\;
        Generate standard \textprob{LPN} samples from the \textprob{Ring-LPN} samples, and then perform Fast Walsh-Hadamard Transform on $l-k$ bits of $\hat{\vect s}$.\;
        Output the $\hat{\vect s}_0$ that maximizes the absolute value of the transform.\;
                \end{algorithm*}
        }

\section{Complexity Analysis}\label{ring-lpn:the-analysis}
%We analyze the attack complexity step by step: 1) The preprocessing step of finding low-weight error position is required for only once; thus, its cost should be omitted. 2) Finding collisions requires $lN + lN^2/2^k$ bit-operations, i.e. the cost of loading samples plus that of performing vector-XOR operation. 3) The last step can be implemented by using Fast Walsh-Hadamard Transformation efficiently, whose cost is $(l-k)2^{l-k}$. Therefore, the following lemma that represents the attacking complexity holds:

Having introduced the two versions in detail, we now estimate their complexity. It is straightforward that the generic algorithm (Algorithm \ref{alg:ring-lpn-attack}) costs $C=C_1+C_2+C_3$ bit-operations. But analyzing the improved attacking complexity for the proposed instance in Lapin is more attractive, which is stated in the following theorem.

\begin{thm}[The complexity of Algorithm \ref{alg:improved-attack}]
  Let $w$ be the weight of the last row in the CRT transformation matrix $\matr{G}_\psi$. Then,
  the complexity of Algorithm, denoted $C^{*}$, is given by
  \begin{equation}\label{ComplexityLemma}
    C^{*} = l\cdot ( N + N^2/2^k)+ (l-k)2^{l-k},
  \end{equation}
  under the condition that $N^2/2^k \geq \frac{1}{\epsilon^{4w}}$.
\end{thm}

\begin{IEEEproof}
    We analyze step by step: 1) First, we store the samples received from $N$ oracle calls into a table using $lN$ bit-operations. 2) Clearing $k$ bits in a collision procedure yields $N^2/2^k$ samples and can be performed in $lN^2/2^k$ bit-operations. As the required standard \textprob{LPN} instance can be read directly from the \textprob{Ring-LPN} instance, the transformation has no computational cost. 3) Afterwards, we perform a Fast Walsh-Hadamard Transform. If the number of unknown bits $l-k$ is at least $\log_2 \p{N^2/2^k}$, then the complexity is $(l-k)2^{l-k}$. This is the final step. Summarizing the individual steps yields $C^*$ which finalizes the proof.
\end{IEEEproof}
%
%\begin{corollary}
%  The complexity of Algorithm , denoted $C_2^{*}$, is given by
%  \begin{equation}\label{ComplexityLemma}
%    C_2^{*} = Nl + 2lN^2/2^k + (l-k)2^{l-k}.
%  \end{equation}
%  under the condition that $lN^2/2^k > \frac{1}{\epsilon^{4w}}$.
%\end{corollary}

\begin{table*}[!t]
        \begin{center}
                \caption{The complexity for attacking each modulus of the proposed instance using reducible polynomial $f$.}\label{table:ring-lpn-parameterandsecurity}
                \begin{tabular}{@{}crrrrrrcrr@{}}\toprule
                        Polynomial $f_i$ & & \multicolumn{3}{c}{Parameters} & & $\log_2 C^*$ \\
                         \cmidrule{2-5}
                        & &\phantom{~~} $k$ & \phantom{~~} $w$   & \phantom{~~} $\log_2 N$  & &  \\
                                \midrule
                         $x^{127}+x^8+x^7+x^3+1$ & &  65 & 26 & 63 && 70.56\\
                         $x^{126}+x^9+x^6+x^5+1$ & &  63 & 26 & 62 && 70.30\\
                         $x^{125}+x^9+x^7+x^4+1$ & &  63 & 26 & 62 && 69.96\\
                         $x^{122}+x^7+x^4+x^3+1$ & &  60 & 27 & 62 && 75.02\\
                         $x^{121}+x^8+x^5+x^1+1$ & &  58 & 29 & 63 && 71.31\\
                        \bottomrule
                \end{tabular}
        \end{center}
\end{table*}

\section{Results}\label{ring-lpn:the-results}
We now present numerical results of the improved partial key recovery attack on the authentication protocol Lapin \cite{Lapin}.
The attack we describe concerns the instance of Lapin using the degree $621$ polynomial $f$
and with the parameter $\eta = \frac{1}{6}$. As claimed in \cite{Lapin}, this given instance is designed to resist the best known attack on \textprob{Ring-LPN} within the complexity $2^{80}$. However, we have shown that it is possible to greatly improve the attack complexity. The improvements can be seen in Table \ref{table:ring-lpn-parameterandsecurity}.

For the distinguishing attack, the complexity is only $2^{69.96}$. For
actual recovery of the secret polynomial, we need all five coordinates
in the CRT representation, which gives an upper bound on the security
that is roughly $2^{75.05}$. However, to solve this search problem, we describe a specified approach presented in the appendix that reduces the complexity to about $2^{71.88}$. A comparison with previous algorithms is given in Table \ref{table:ring-lpn-comparison} in the introduction.

\section{Concluding Remarks}\label{ring-lpn:the-conclusion}
We have proposed a new generic algorithm to solve the reducible case of
\textprob{Ring-LPN}. By exploiting the ring structure further, our new
algorithm is much more efficient than previous algorithms,
enough to break the claimed 80-bit security for one of the two proposed instances of Lapin.

We have shown that a linear code arising from the CRT transform
characterizes the performance of this attack through its minimum
distance. This is combined with some standard techniques of using
birthday or possibly generalized birthday arguments and
efficient recovery through Fast Walsh-Hadamard transform.

The low-weight property of the polynomials in the Lapin case makes the
problem considerably easier than otherwise and thus makes Lapin susceptible to our attack.
Using really low-weight irreducible polynomials such as $x^{127}+x+1$ can give
rise to linear relations with weight as low as 10 or even less. We
have not seen that such polynomials have been pointed out as very weak before.

The description of the new algorithm was influenced by the Lapin
case. There are more improvements that can be described in the
general case. One such improvement is the use of a generalized
birthday technique \cite{Wagner}. This will allow us to consider
larger dimensions at the cost of increasing the noise level. We have
also noted that the simple bit-oriented samples in this paper can be
replaced by more complicated vectorial samples, which will give a
stronger bias.

\appendices

\section{Lapin Authentication Protocol}
In this section, we describe the Lapin two-round authentication protocol. Let $R = \gf{2}{}[x]/(f)$ be a ring and $R^*$ the set of units in $R$. The protocol is defined over the ring $R$. Let $\pi$ be a mapping, chosen such that for all $c,c'\in \{0,1\}^\lambda$, $\pi(c) - \pi(c') \in R \setminus R^*$ if and only if $c=c'$. Furthermore, let $\eta \in \p{0,\frac{1}{2}}$ be a Bernoulli distribution parameter and $\eta' \in \p{\eta,\frac{1}{2}}$ a threshold parameter. The elements $R$, $\pi:\{0,1\}^\lambda \rightarrow R$, $\eta$ and $\eta'$ are public parameters. The ring elements $s,s' \in R$ constitute the secret key.

\protocolwrap{Lapin two-round authentication}{prot:ring-lpn-lapin}{
$\begin{array}{@{}l@{}l@{}c@{}l@{}}
&\textnormal{\bf{Tag}} && \textnormal{\bf{Reader}} \\
&\hspace{2.5cm}&\hspace{1.5cm}&c \setrandl \{0,1\}^\lambda \\
&&\xleftarrow{\phantom{~~~}\textstyle c\phantom{~~~}}&\hspace{3.6cm}\\
&r \setrandl R^*; e \setrandl\distr{Ber}_\eta^R&  \\
&z \leftarrow r\cdot (s \cdot \pi(c) + s') + e &  \\
&& \xrightarrow {\phantom{~}\textstyle (r,z)\phantom{~}} \\
&&&\textnormal{if $r \not \in R^*$ then \textsf{reject}} \\
&&& e' \leftarrow z - r \cdot (s \cdot \pi(c) + s') \\
&&&\textnormal{if $\hamw{e'} > n\cdot \eta'$ then \textsf{reject}} \\
&&& \textnormal{else \textsf{accept}}
\end{array}$
}
Suppose that we have a key-generation oracle; it will give us the key $s,s' \setrandl R$. The secret key is shared among the tag and the reader. Protocol \ref{prot:ring-lpn-lapin} gives how information is exchanged between the tag and the reader in Lapin.

\section{A Special Secret Recovery Approach}
In this section, we describe a better secret recovering approach on the instance proposed in Lapin with a reducible polynomial. This attack exploits the secret polynomial's representations in the three relatively easy-attacked quotient rings $\gf{2}{}[x]/(f_i), i = 1, 2, 3$, to decrypt those in rings $\gf{2}{}[x]/(f_i), i = 4, 5$, and thus obtains higher efficiency than simply attacking one ring by another. Actually, it reduces the attacking complexity from $2^{75.05}$ to $2^{71.88}$.

Denote $\vect a|| \vect b$ as the concatenation of two vectors $\vect a$ and $\vect b$, and $(s_1 \bmod {f_1}, s_2 \bmod {f_2},\ldots , s_5 \bmod{f_5})$ as the CRT representation of $s$. The three-step attack is described as follows.

We first recover the secret polynomials $s_1$, $s_2$ and $s_3$ in the corresponding smaller rings. This step costs around $2^{71.88}$ bit operations. Then, using these known polynomials, we show that the secrets $s_4$ and $s_5$ can be recovered with negligible costs.

The second step is transforming each \textprob{Ring-LPN} sample to its corresponding standard \textprob{LPN} samples. This step will be tricky as we want to make use of the known information. Since $f_i$ are distinct irreducible polynomials over $\gf{2}{}[x]$, there exist polynomials $t_i \in R$ such that $s = \sum^{5}_{i=1} s_i \cdot t_i$ and $t_i \equiv 1 \bmod {f_i}$, where $i = 1, 2, \ldots ,5$. Moreover, these polynomials $t_i, i = 1, 2, \ldots ,5$ can be computed efficiently. Then, for each sample ($r, v$), where $v = r \cdot s + e$, by adding the polynomial $r \cdot (\sum_{i=1}^3s_i \cdot t_i)$ to $v$, we create a new polynomial $r \cdot (s_4 \cdot t_4 + s_5 \cdot t_5) + e$,  which will be converted to a vector $\matr A_4  \vect s_4 + \matr A_5  \vect s_5 + \vect e$, where $\matr A_4(\matr A_5)$ is the matrix whose $i$-th column is the coefficient vector of $r \cdot t_4 \cdot x^i (r \cdot t_5 \cdot x^i) \bmod f$. The degree of $s_4$($s_5$) is $122$($121$), thereby yielding that only the first $122$($121$) columns of the matrix $A_4(A_5)$, denoted $A'_4(A'_5)$, are useful. We can compute those sub-matrices in $2^{17.2}$ bit operations for each \textprob{Ring-LPN} sample, thereby economically constructing $621$ standard \textprob{LPN} samples $(\vect r^i_4||\vect r^i_5, \scalar{\vect s_4||\vect s_5}{\vect r^i_4||\vect r^i_5} + e_i)$ for $i = 0, 1, \ldots , 620$, where $\vect r^i_4 (\vect r^i_5)$ is the $i$-th row in the matrix $\matr A'_4(\matr A'_5)$.

The last step is to attack a standard \textprob{LPN} problem with length $243$ and error probability $1/6$. By Levieil-Fouque algorithm, the cost is small, i.e., around $2^{57.1}$ bit operations. It requires $2^{46.6}$ standard \textprob{LPN} samples, i.e. $2^{37.3}$ \textprob{Ring-LPN} ones.

The number of samples in the second step is also bounded by that required in the final step; the overall cost of the last two steps, therefore, is no more than $2^{60}$ bit operations, far less than that required in the starting step. Thus, it is reasonable to embrace $2^{71.88}$ bit operations  as the total attacking complexity.

%\bibliographystyle{plain}
%\bibliography{U:/martin/bibtex/cryptogroup}

\begin{thebibliography}{99}

\bibitem{BFKL}
Blum, A., Furst, M., Kearns, M., Lipton, R.: Cryptographic Primitives Based on Hard Learning Problems. In: Stinson, R. (ed.) CRYPTO 1993, LNCS, vol. 773, pp. 278--291. Springer, Heidelberg (1994)

\bibitem{BKW}
Blum, A., Kalai, A., Wasserman, H.: Noise-Tolerant Learning, the Parity Problem, and the Statistical Query Model. In: Journal of the ACM, vol. 50, no. 4, pp. 506--519. (2003)

\bibitem{NP}
Berlekamp, E.R., McEliece, R.J., van Tilborg, H.C.A.: On the Inherent Intractability of Certain Coding Problems. IEEE Trans. Info. Theory, vol. 24, pp. 384--386. (1978)

\bibitem{Bernstein}
Bernstein, D., Lange T.: Never trust a bunny. In: Radio Frequency Identification Security and Privacy Issues, pp. 137--148. Springer, Berlin Heidelberg (2013)

\bibitem{FastCorrelationAttackAlgorithmicView}
Chose, P., Joux, A., Mitton, M.: Fast Correlation Attacks: An Algorithmic Point of View. In: Knudsen, L.R. (ed.) EUROCRYPT 2002. LNCS, vol. 2332, pp. 209--221. Springer, Heidelberg (2002)

\bibitem{Damgard}
Damgard, I., Park, S.: Is Public-Key Encryption Based on LPN Practical? Cryptology ePrint Archive, Report 2012/699
(2012), http://eprint.iacr.org/

\bibitem{Dodis}
Dodis, Y., Kiltz, E., Pietrzak, K., Wichs, D.: Message Authentication, Revisited. In: Pointcheval, D., Johansson, T. (eds.) EUROCRYPT 2012. LNCS, vol. 7237, pp. 355--374. Springer, Heidelberg (2012)

\bibitem{Imai}
Fossorier, M.P.C., Mihaljevic, M.J., Imai, H., Cui, Y., Matsuura, K.: A Novel Algorithm for Solving the LPN Problem and its Application to Security Evaluation of the HB Protocol for RFID Authentication. Cryptology ePrint archive, Report 2012/197 (2012), http://eprint.iacr.org/

\bibitem{Mask-Lapin}
Gaspar, L., Leurent, G., Standaert, F. X.: Hardware Implementation and Side-Channel Analysis of Lapin. CT-RSA 2014, (2014)

\bibitem{Gilbert08}
Gilbert, H., Robshaw, M.J.B., Seurin, Y.: HB$^\#$: Increasing the Security and the Efficiency of HB$^+$. In: Smart, N.P. (ed.) EUROCRYPT 2008. LNCS, vol. 4965, pp. 361--378. Springer, Heidelberg (2008)

\bibitem{Gilbert05}
Gilbert, H., Robshaw, M.J.B., Sibert, H.: An active attack against HB$^+$---a provably
secure lightweight authentication protocol. Cryptology ePrint Archive, Report
2005/237 (2005), http://eprint.iacr.org/

\bibitem{Lapin}
Heyse, S., Kiltz, E., Lyubashevsky, V., Paar, C., Pietrzak, K.: Lapin: An Efficient Authentication Protocol Based on Ring-LPN. In: FSE 2012, pp. 346--365. (2012)

\bibitem{HB}
Hopper, N.J., Blum, M.: Secure human identification protocols. In: Boyd, C. (ed.) ASIACRYPT 2001. LNCS, vol. 2248, pp. 52--66. Springer, Heidelberg (2001)

\bibitem{JW}
Juels, A., Weis, S.A.: Authenticating pervasive devices with human protocols.
In: Shoup, V. (ed.) CRYPTO 2005. LNCS, vol. 3621, pp. 293--308. Springer,
Heidelberg (2005)

\bibitem{KatzShin}
Katz, J., Shin, J.S.: Parallel and concurrent security of the HB and HB$^+$ protocols.
In: Vaudenay, S. (ed.) EUROCRYPT 2006. LNCS, vol. 4004, pp. 73--87. Springer,
Heidelberg (2006)

\bibitem{Katz}
Katz, J., Shin, J.S., Smith, A.: Parallel and concurrent security of the HB and
HB$^+$ protocols. Journal of Cryptology 23(3), 402--421 (2010)

\bibitem{Kearns}
Kearns., M: Effcient Noise-Tolerant Learning from Statistical Queries. In: J. ACM 45(6), pp. 983--1006. (1998)

\bibitem{Klitz}
Kiltz, E., Pietrzak, K., Cash, D., Jain, A., Venturi, D.: Efficient Authentication from Hard Learning Problems. In: Paterson, K.G. (ed.) EUROCRYPT 2011. LNCS, vol. 6632, pp. 7--26. Springer, Heidelberg (2011)

\bibitem{Kirchner}
Kirchner, P.: Improved Generalized Birthday Attack. Cryptology ePrint Archive, Report 2011/377
(2011), http://eprint.iacr.org/

\bibitem{LF06}
Levieil, E., Fouque, P. A.: An Improved LPN Algorithm. In: Proceedings of SCN 2006, LNCS 4116, pp. 348--359. Springer, Heidelberg (2006)

\bibitem{Lyubashevsky}
Lyubashevsky, V.: The Parity Problem in the Presence of Noise, Decoding Random Linear Codes, and the Subset Sum Problem, In: Chekuri, C., Jansen, K., Rolim, J.D.P., Trevisan, L. (eds.)  APPROX-RANDOM 2005, LNCS, vol. 3624, pp. 378--389. Springer, Heidelberg (2005)

\bibitem{HB-MP}
Munilla, J., Peinado, A.: HB-MP: A further step in the HB-family of lightweight
authentication protocols. Computer Networks 51(9), pp. 2262--2267. (2007)

\bibitem{Regev}
Regev, O.: On Lattices, Learning with Errors, Random Linear Codes, and Cryptography. In: Gabow, H.N., Fagin, R. (eds.) 37th Annual ACM Symposium on Theory of Computing, Proceedings, pp. 84--93. (2005)

\bibitem{Stern89}
Stern, J.: A Method for Finding Codewords of Small Weight. In: Wolfmann, J., Cohen, G. (eds.) Coding Theory 1988. LNCS, vol. 388, pp. 106--113. Springer, Heidelberg (1989)

\bibitem{Stern}
Stern, J.: A New Identification Scheme Based on Syndrome Decoding. In: Stinson,
D.R. (ed.) CRYPTO 1993. LNCS, vol. 773, pp. 13--21. Springer, Heidelberg (1994)

\bibitem{Wagner}
Wagner, D.: A Generalized Birthday Problem. In: Yung, M. (ed.) CRYPTO 2002. LNCS, vol. 2442, pp. 288--304. Springer, Heidelberg (2002)

\end{thebibliography}


%\end{IEEEbiography}
%% insert where needed to balance the two columns on the last page with
%%\newpage


%\vfill


\end{document}